\documentclass{aip-cp}

\usepackage[numbers]{natbib}
\usepackage{rotating}
\usepackage{graphicx}
\usepackage{bm}
\usepackage[utf8]{inputenc}

\newcommand{\vect}[1]{{\boldsymbol{#1}}}

\begin{document}

% Title portion
\title{Two-dimensional Hybrid Simulations of Kinetic Plasma
Turbulence: Current and Vorticity vs Proton Temperature}

\author[aff1,aff2]{Luca Franci}
\author[aff3]{Petr Hellinger\corref{cor1}}
\author[aff4]{Lorenzo Matteini}
\author[aff1,aff5]{Andrea Verdini}
\author[aff1,aff6]{Simone Landi}

\affil[aff1]{Dipartimento di Fisica e Astronomia, Università degli Studi di Firenze, Largo E. Fermi 2, I-50125 Firenze, Italy}
\affil[aff2]{INFN-Sezione di Firenze, Via G. Sansone 1, I-50019 Sesto F.no (Firenze), Italy}
\affil[aff3]{Astronomical Institute, AS CR, Bocni II/1401, CZ-14100 Prague, Czech Republic}
\affil[aff4]{Physics Department, Imperial College London, London SW7 2AZ, UK}
\affil[aff5]{Solar-Terrestrial Center of Excellence, Royal Observatory of Belgium, Brussels, Belgium}
\affil[aff6]{INAF-Osservatorio Astrofisico di Arcetri, Largo E. Fermi 5, I-50125 Firenze, Italy}
\corresp[cor1]{Corresponding author: petr.hellinger@asu.cas.cz}

\maketitle

\begin{abstract}
Proton temperature anisotropies between the directions parallel and
perpendicular to the mean magnetic field are usually observed in the
solar wind plasma. Here, we employ a high-resolution hybrid
particle-in-cell simulation in
order to investigate the relation between spatial properties of
the proton temperature and the peaks in the current density and in the
flow vorticity.
Our results indicate that, although regions where the proton
temperature is enhanced and temperature anisotropies are larger
correspond approximately to regions where many thin current sheets
form, no firm quantitative evidence supports the
idea of a direct causality between the two phenomena.  On the
other hand, quite a clear correlation between the behavior of the proton
temperature and the out-of-plane vorticity is obtained.
\end{abstract}

\section{INTRODUCTION}

The solar wind is a highly turbulent plasma. This idea is supported
by the power-law behavior of its observed energy spectra, which span
nearly four decades in frequency, from large to small kinetic scales
(e.g., \citep{Alexandrova_al_2013}). Among other things, in situ
measurements also reveal the presence of an ubiquitous proton
temperature anisotropy between the direction parallel and
perpendicular to the mean magnetic field \citep{Marsch_al_1982,
  Hellinger_al_2006}.  Vlasov-Hybrid simulations suggest that such
temperature anisotropy and non-Maxwellian kinetic effects are mostly
found around peaks of the current density
\citep{Greco_al_2012, Servidio_al_2012, Valentini_al_2014, Servidio_al_2015}.

Recently, high-resolution two-dimensional (2D) hybrid particle-in-cell
simulations have proved to be a reliable, state-of-the-art tool to
investigate the properties of kinetic plasma turbulence, provided that
a sufficiently large number of particles-per-cell is employed,
especially when trying to quantitatively estimate the perpendicular
proton temperature. In particular, the direct numerical simulations
shown in \citep{Franci_al_2015a, Franci_al_2015b} have been able to
recover simultaneously several features observed in the solar wind
spectra, e.g.: i) a power-law behavior for the magnetic, kinetic and
residual energy spectra with different spectral indices (e.g.,
\citep{Chen_al_2013}), ii) a magnetic spectrum with a smooth break at
proton scales and a power-law scaling in the sub-proton range with a
spectral index of $\sim-2.8$ (e.g., \citep{Sahraoui_al_2010}), iii) an
increase in magnetic compressibility at small scales (e.g.,
\citep{Kiyani_al_2013}), iv) a strong coupling between density and
magnetic fluctuations in the kinetic range (e.g.,
\citep{Chen_al_2013}).

In the present paper, we show new complementary results coming from
the 2D hybrid particle-in-cell simulations already presented in
\citep{Franci_al_2015a, Franci_al_2015b}.  In particular, we will
focus our attention on the correlations between the peaks in the
out-of-plane vorticity and the proton temperature enhancement and
anisotropy.

\section{RESULTS}

The numerical results discussed here were obtained by means of the
hybrid particle-in-cell code CAMELIA, which treats electrons as a
massless, charge neutralizing, isothermal fluid, whereas ions as
particles. The characteristic spatial unit is the proton inertial
length, $d_p$. We employ a 2D square computational grid in the
$(x,\,y)$ plane, with periodic boundary conditions, $2048^2$ square
cells and a total length $L = 256 \, d_p$. Each cell has a size of
$0.125 \, d_p$ and contains $8000$ particles representing protons. The
number density is assumed to be equal for protons and electrons and
both species are isotropic, with the same plasma beta, $\beta=0.5$.
The initial proton temperature anisotropy is set to $A_{p} =
T_{p\perp} / T_{p\parallel} = 1$, where $T_{p\perp}$ and
$T_{p\parallel}$ are the perpendicular and parallel proton
temperatures, respectively.  We impose an initial ambient magnetic
field $\vect{B}_0 = B_0 \hat{\vect{z}}$, perpendicular to the
simulation plane, and an initial spectrum of linearly polarized
magnetic and bulk velocity fluctuations with only in-plane components.
Fourier modes of equal amplitude and random phases are excited,
assuring energy equipartition and vanishing correlation between
kinetic and magnetic fluctuations.  Fields are defined as parallel
($\parallel$) and perpendicular ($\perp$) with respect to the
\textit{mean} magnetic field, whereas the proton
temperatures are intended with respect to
the \textit{local} magnetic field, $\vect{B} = \vect{B}_0 + \delta
\vect{B}$, where $\delta \vect{B}$ are the fluctuations.  For further
information about the numerical setting and parameters and for a
complete definition of all quantities, please refer to
\citep{Franci_al_2015a, Franci_al_2015b}.
\begin{figure}[hb]
  \begin{tabular}{ccc}
    \includegraphics[width=0.38\textwidth,trim={0 0 0
        1cm},clip=true]{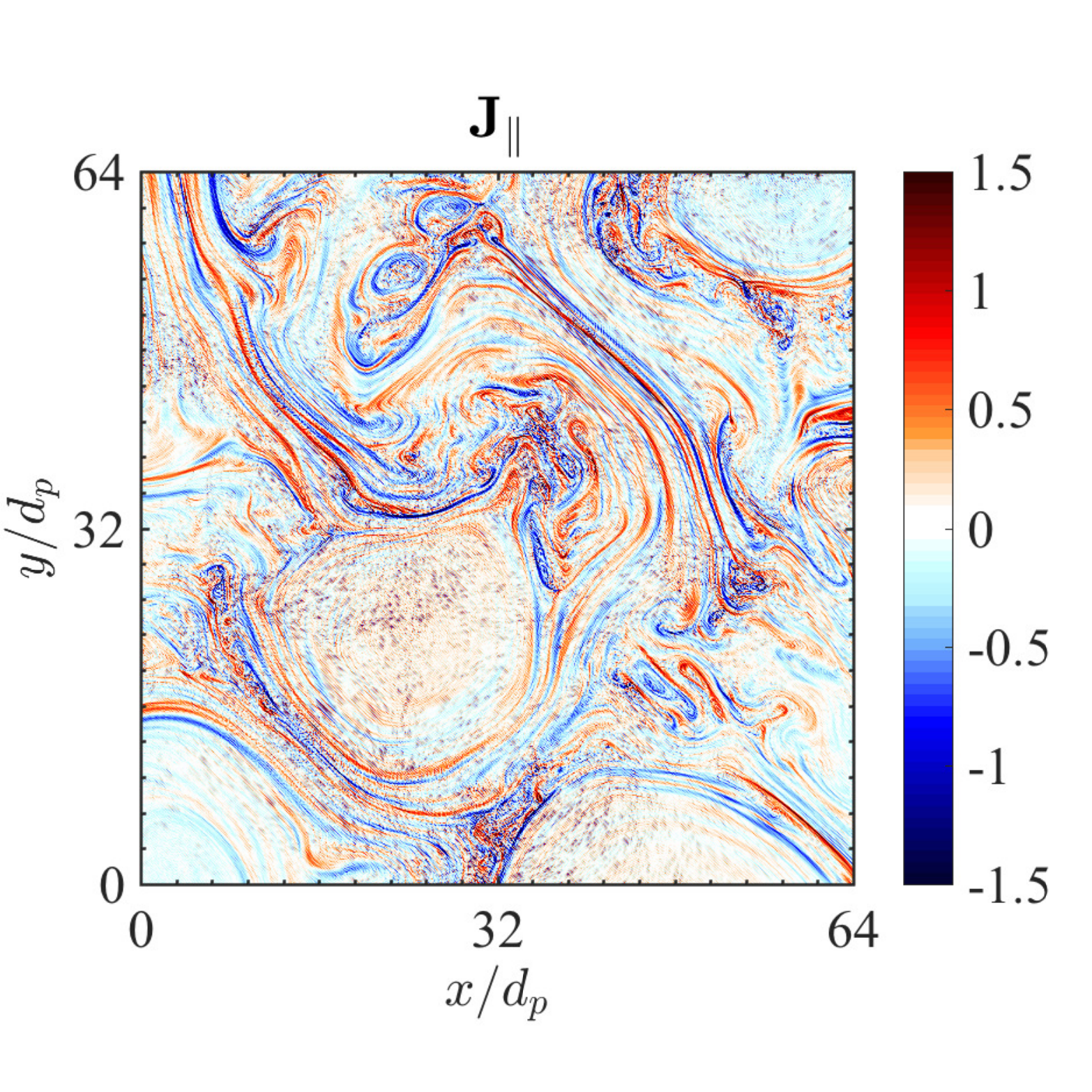}
    \includegraphics[width=0.38\textwidth,trim={0 0 0
        1cm},clip=true]{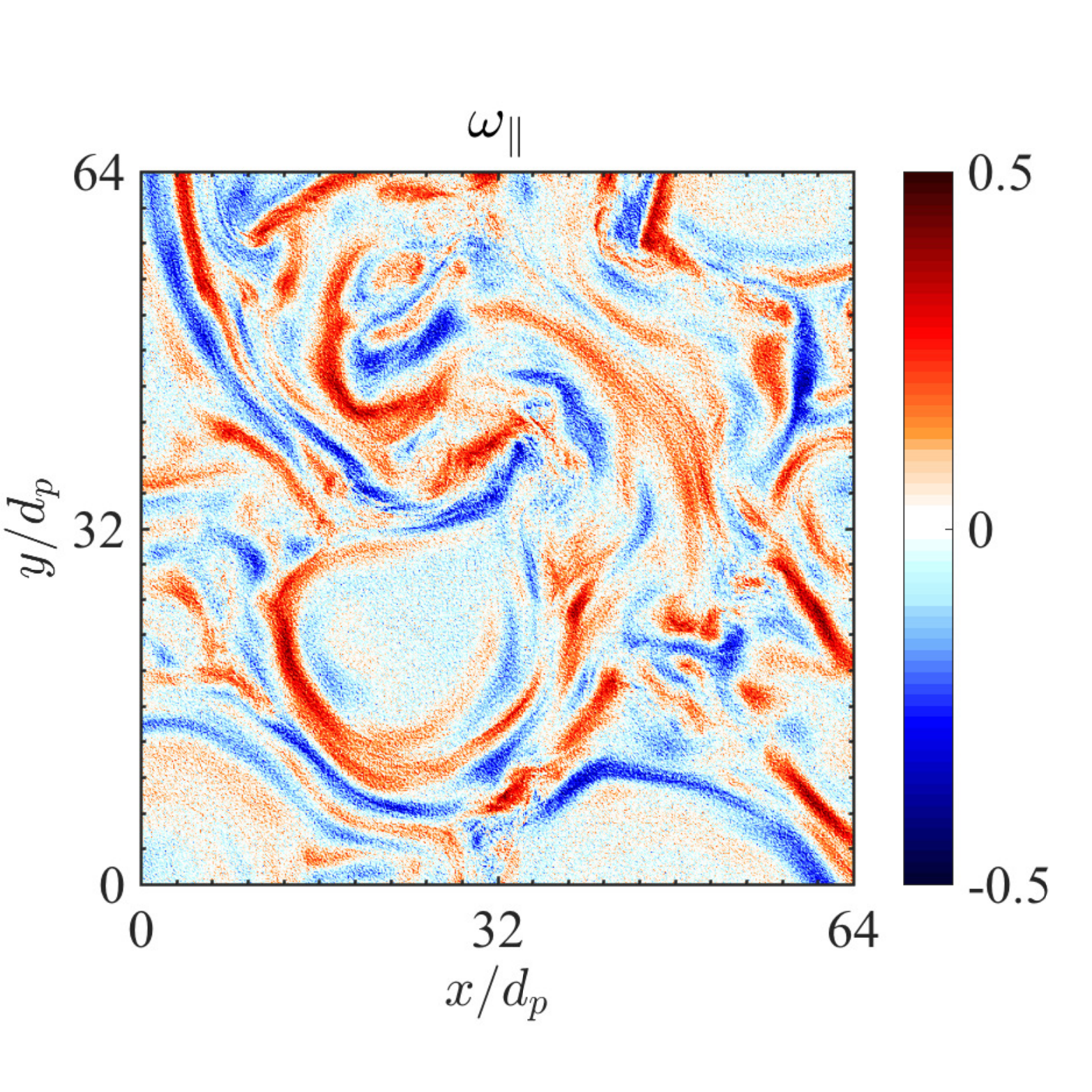}
    \vspace{-0.5cm} \\ 
    \includegraphics[width=0.38\textwidth,trim={0 0
        0 1cm},clip=true]{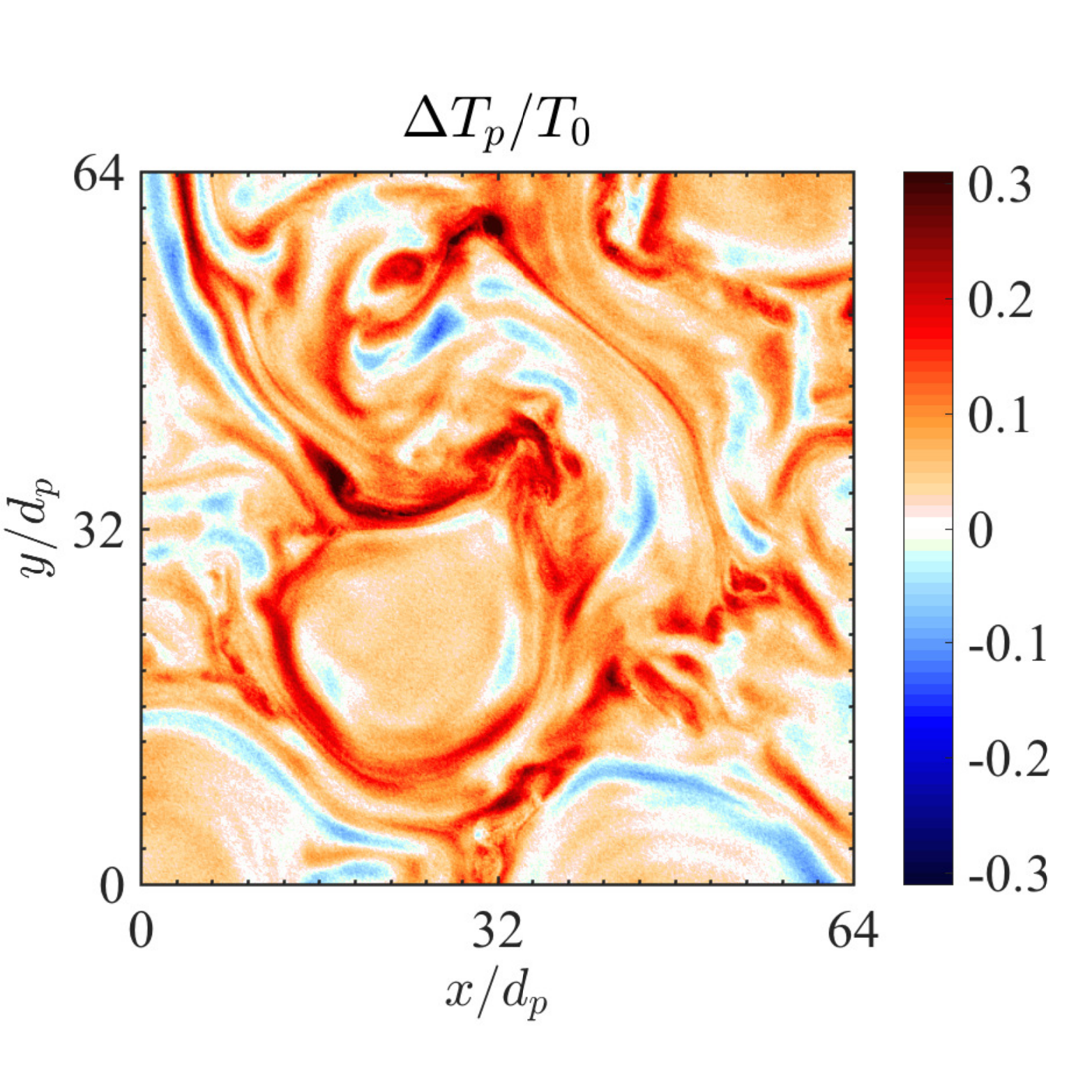}
    \includegraphics[width=0.38\textwidth,trim={0 0 0
        1cm},clip=true]{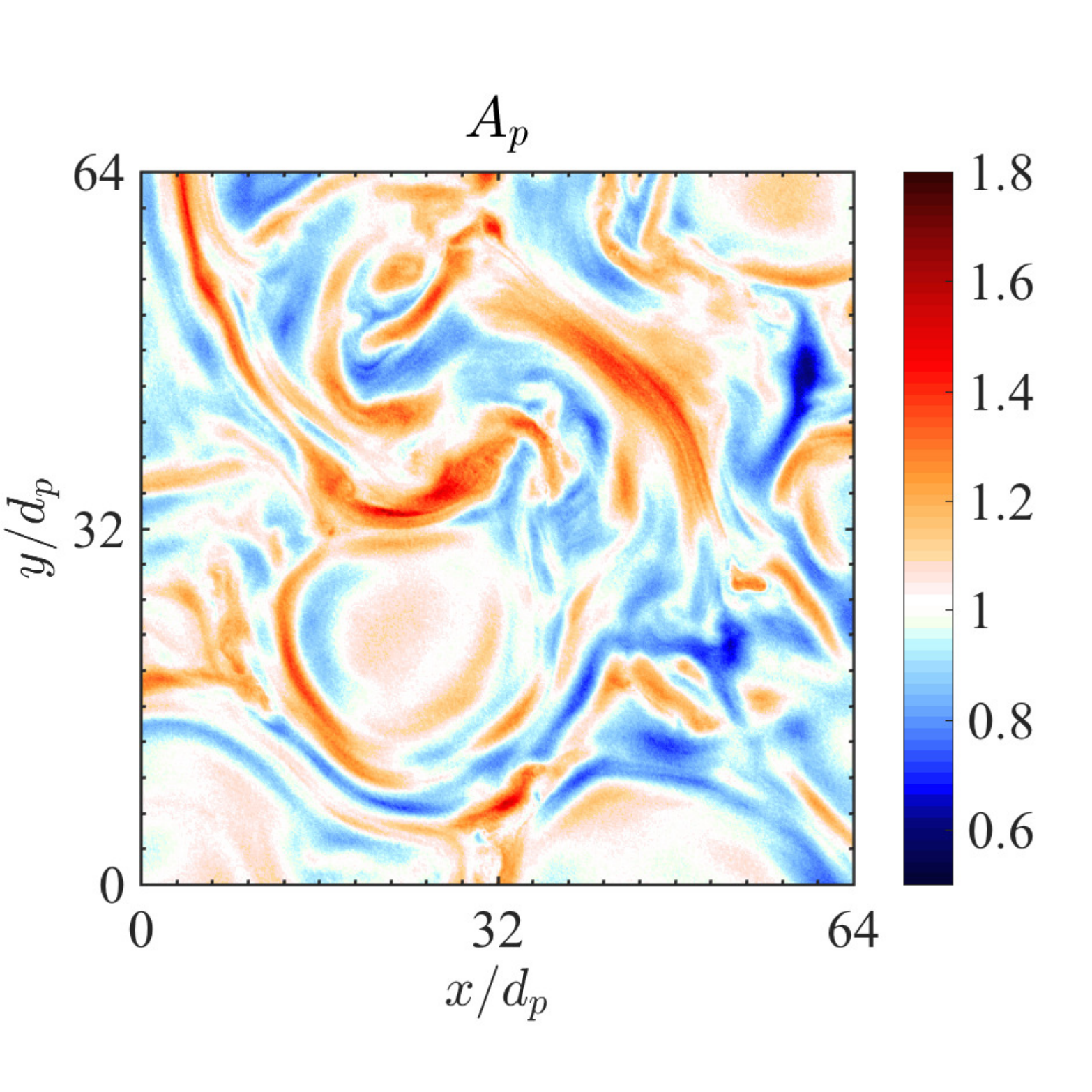}
  \end{tabular}
  \caption{Contour plots of four different quantities on a portion of
    the $(x,y)$ plane, i.e., $[0,64\,d_p]\times[0,64\,d_p]$, at 
    the time of maximum turbulent activity: the out-of-plane
    current density, $\vect{J}_\parallel$ (\textit{top-left panel}),
    the out-of-plane vorticity, $\vect{\omega}_\parallel$
    (\textit{top-right panel}), the proton temperature variation
    normalized to the initial temperature, $\Delta T_p/T_0$
    (\textit{bottom-left panel}), and the proton temperature
    anisotropy, $A_{p}$ (\textit{bottom-right panel}).}
  \label{fig:isocontours}
\end{figure} 

Figure~\ref{fig:isocontours} shows isocontours of four different
quantities in a small portion of the 2D simulation domain, i.e.,
$[0,\,64\,d_p] \times [0,\,64\,d_p]$.  All these snapshots have been
taken at the time of maximum turbulent activity, which corresponds to
the peak of the RMS value of the out-of-plane current density,
$J_\parallel = (\nabla \times \vect{B})_\parallel$, during its time
evolution \citep{Franci_al_2015a, Franci_al_2015b}.  In the top-left
panel, we show the local spatial distribution of the out-of-plane
current density, $J_\parallel$. By the time a turbulence cascade has
fully developed, many thin current sheets have already formed and
partially disrupted, generating a complex pattern with small-scale
structures.  In the top-right panel, we report the out-of-plane
vorticity, $\vect{\omega}_\parallel = (\nabla \times
\vect{u})_\parallel$, where $\vect{u}$ is the proton bulk velocity.
It seems to follow a similar pattern as the one of
$\vect{J}_\parallel$, although with a much less filamentary structure.
Peaks of $\vect{\omega}_\parallel$ and peaks of $\vect{J}_\parallel$
occupy approximately the same regions, although the latter exhibits a
more structured pattern and it usually fills the spaces between the
structures of the former. In the bottom-left panel, we report the
normalized proton temperature variation, $\Delta T_p/T_0 = (T_p -
T_0)/T_0$, where $T_p = (2\, T_{p\perp} + T_{p\parallel})/3$ is the
average proton temperature at the time of maximum turbulent activity.
Although $\Delta T_p$ can be locally both negative or positive, the
resulting global proton temperature enhancement is clearly positive,
and the same holds when the whole domain is considered
(cf.~\citep{Franci_al_2015b}). Finally, the proton temperature
anisotropy, $A_p$, is reported in the bottom-right panel.  It ranges
about between 0.6 and 1.6 in this portion of the computational
domain (a similar range of values is reached in the
whole box). This wide excursion is a signature of a strong local
reshaping of particle distributions, leading to both perpendicular and
parallel anisotropies \citep{Servidio_al_2014}.

If we now compare the local spatial distribution of these four
quantities, we see that proton temperature enhancements and a quite
strong proton temperature anisotropy seem to occur in the vicinity
of current sheets (cf.,~\citep{Servidio_al_2012}).  Nevertheless, if
we now focus on the structure of the out-of-plane vorticity, we
realize that it matches the shapes of the two quantities related to
the proton temperatures even better. Moreover, areas with a positive
vorticity usually correspond to an increase in the total proton
temperature and to the proton temperature anisotropy larger than 1
($T_{p\perp} > T_{p\parallel}$), whereas the opposite behavior occurs
in areas with a negative vorticity.
\begin{figure}[ht]
  \begin{tabular}{ccc}
    \includegraphics[width=0.37\textwidth]{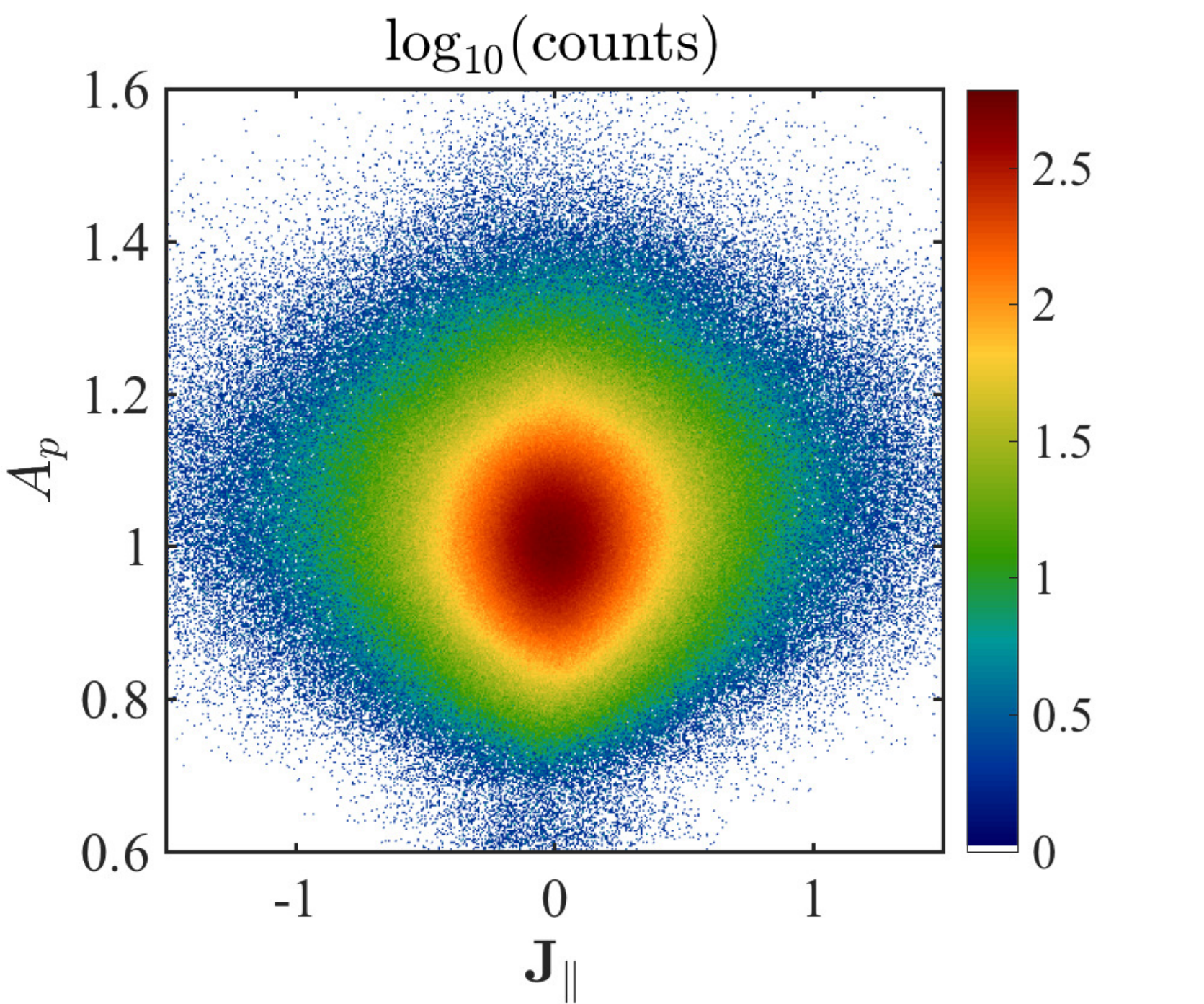}
    \includegraphics[width=0.37\textwidth]{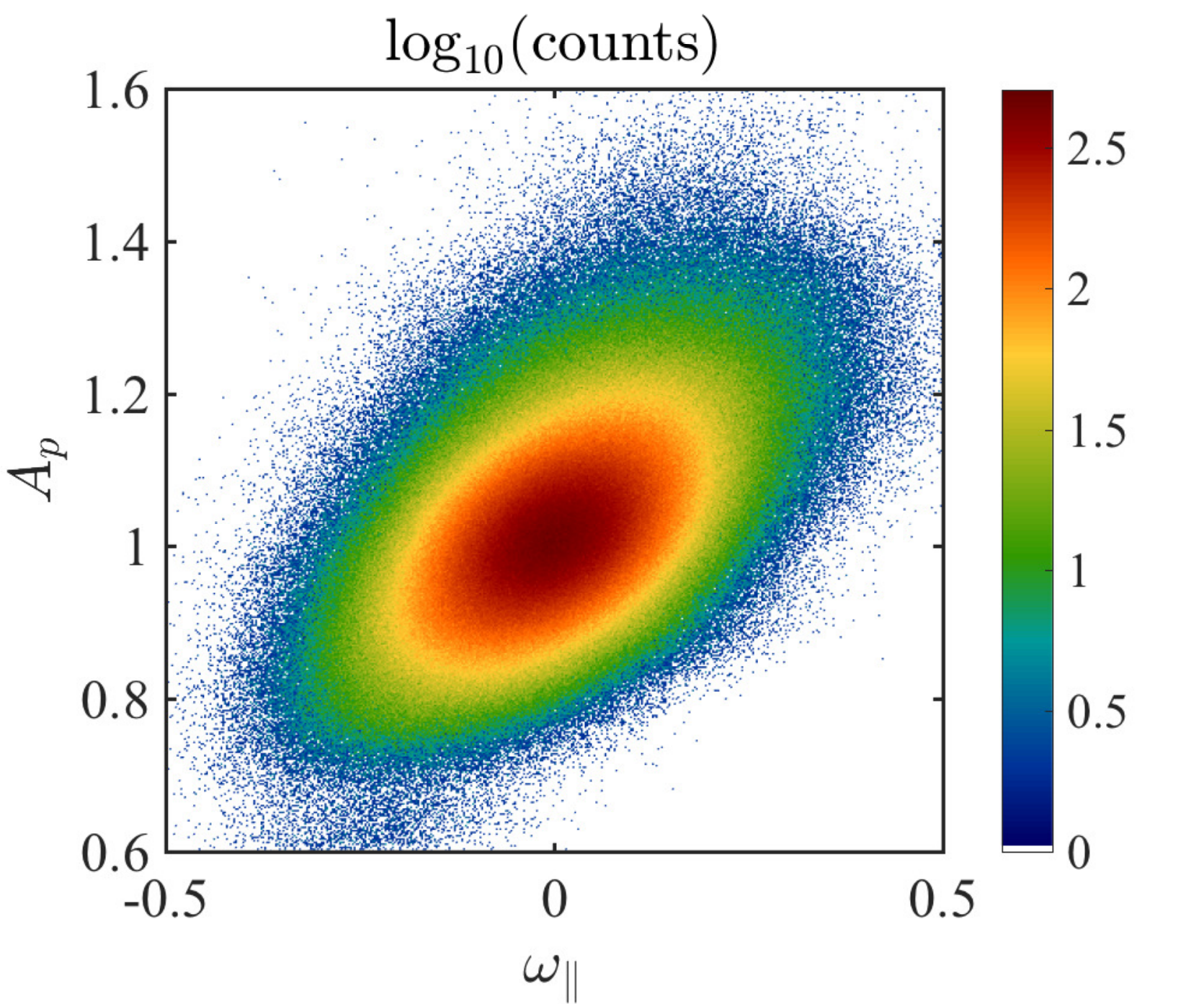} \\ 
    \includegraphics[width=0.37\textwidth]{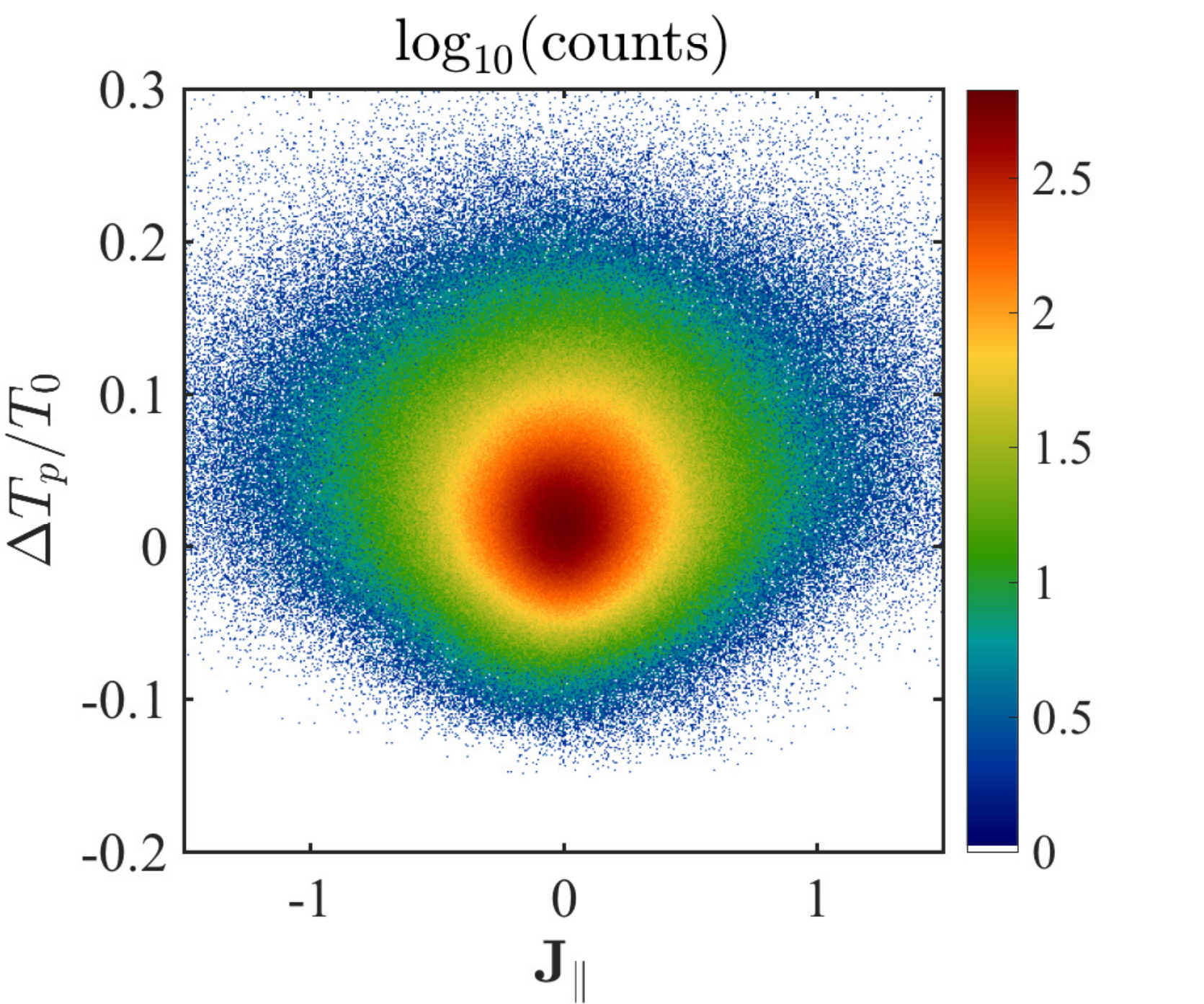}
    \includegraphics[width=0.37\textwidth]{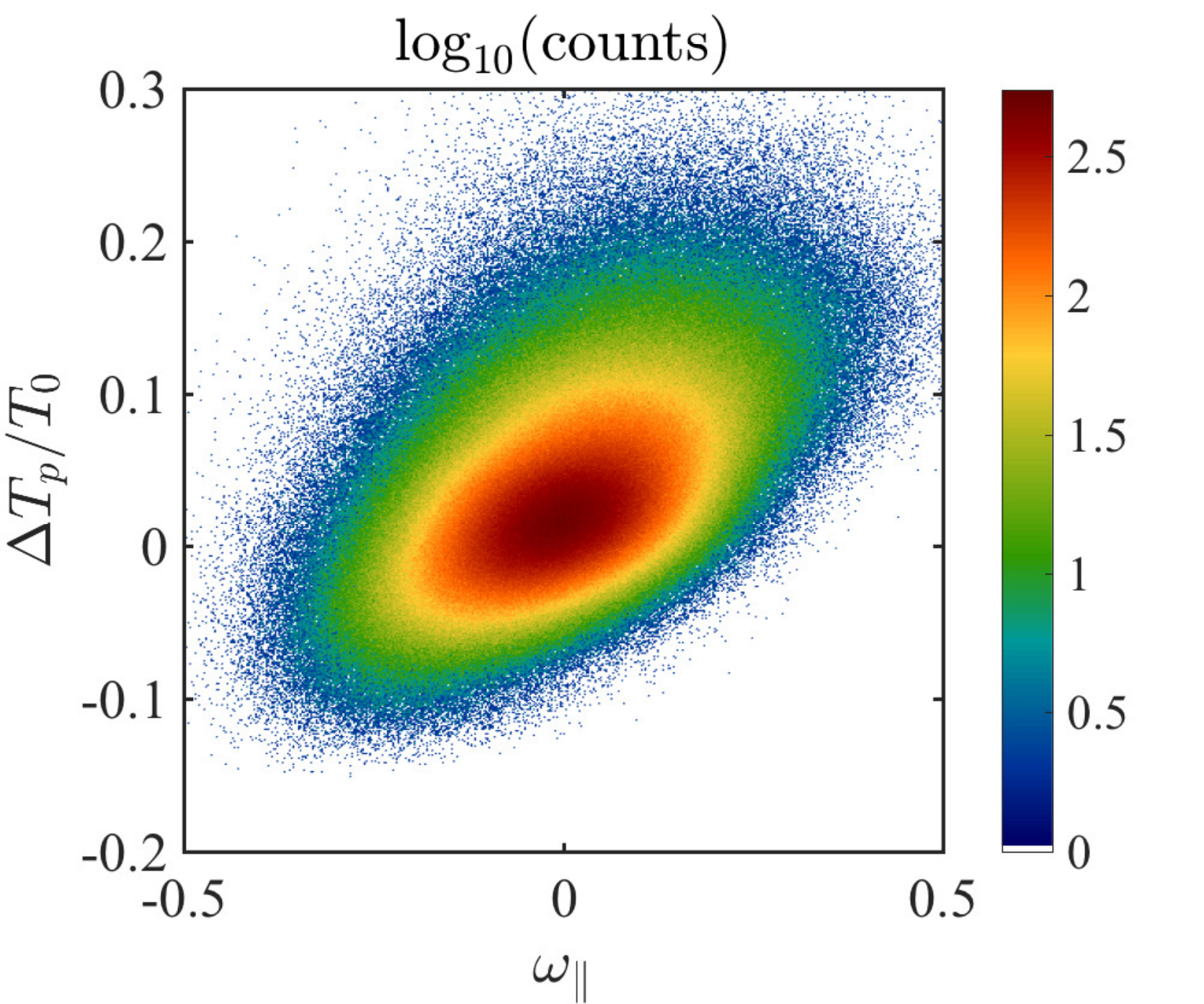}
  \end{tabular}
  \caption{Correlations among the four different quantities
    shown in Fig.~\ref{fig:isocontours}, i.e., the out-of-plane
    current density, $\vect{J}_\parallel$, the out-of-plane vorticity,
    $\vect{\omega}_\parallel$, the relative proton temperature variation, 
    $\Delta T_p/T_0$, and the proton temperature anisotropy, $A_{p}$.}
    \label{fig:correlations}
\end{figure} 

In order to go beyond a qualitative estimate of the correlation among
the four above-mentioned quantities, in Fig.~\ref{fig:correlations} we
show four scatter plots. In particular, in the top-left and in the
bottom-left panels we report the values of the proton temperature
anisotropy, $A_{p}$, and of the relative proton temperature variation,
$\Delta T_p/T_0$, respectively, as a function of 
the out-of-plane current density, $\vect{J}_\parallel$.  No clear
correlation with $\vect{J}_\parallel$ can be appreciated for both
quantities. In the top-right and in the bottom-right panels we show
the value of $A_{p}$ and $\Delta T_p/T_0$ as a function of the
out-of-plane vorticity, $\vect{\omega}_\parallel$. In
this case, a correlation with $\vect{\omega}_\parallel$ is clearly
present for both quantities, confirming what we have already inferred
when looking at the respective isocontour plots.

\section{DISCUSSION}

We have shown that regions with a large proton temperature anisotropy
and with a global proton heating seem to be more directly correlated
with the structures in the vorticity, rather than with the current
density.  The local spatial distribution of the latter exhibits a
similar pattern, but it has a more complex and filamentary structure,
as expected from the spectra of the fluctuations (see
\citep{Franci_al_2015a, Franci_al_2015b}), since the magnetic field
has much more power than the proton bulk velocity at small scales.  A
possible explanation for the direct link between the vorticity and the
proton temperature could be related to the development of
non-gyrotropic ion distribution functions. The presence of such
non-gyrotropic distributions has been observed in hybrid-Vlasov
simulations \citep{Valentini_al_2014}.  Recently, it has been shown
that a sheared velocity field can provide an effective mechanism that
makes an initial isotropic state anisotropic and non-gyrotropic
\citep{DelSarto_al_2015}. The equations for the evolution of the
diagonal terms of the pressure tensor involve the cross-derivatives of
the flow velocity components. This could determine the observed
correlation between the evolution of the proton temperature and the
vorticity. In particular, according to Eq.~(2) of
\citep{DelSarto_al_2015}, the evolution of the 2nd order anisotropic ion
moment is related to the sign of the scalar product 
$\mathbf{B} \cdot \mathbf{\omega}$ and therefore, in our case, to the
sign of the vorticity alone, since the global magnetic field is always
directed along the positive $z$ direction. This is consistent with
what we observe, since positive and negative vorticity (and not only
its intensity) differently controls the variation in the proton
temperatures. In particular, the vorticity is found to be
correlated/anticorrelated with the perpendicular/parallel proton
temperature, since regions with the positive vorticity exhibit an increase in
$T_{p\perp}$ and a decrease in $T_{p\parallel}$, while the opposite occurs for the negative
vorticity.  This explains the increase of both the total proton temperature 
and the proton temperature anisotropy in regions with the positive vorticity.
 The observed correlations are likely related to a more
general relationship between the velocity stress tensor and the
pressure tensor. Further work is therefore needed to better clarify the
effects of the former (and in particular of the
vorticity) on the latter.

% Acknowledgement
\section{ACKNOWLEDGMENTS}
The authors wish to acknowledge valuable discussions with Daniele Del Sarto.
This project has received funding from the European Union’s Seventh
Framework Programme for research, technological development and
demonstration under grant agreement no 284515 (SHOCK). Website:
http://project-shock.eu/home/.  This research was conducted with high
performance computing (HPC) resources provided by CINECA (grant HP10CVCUF1).
L.M. was funded by STFC grant ST/K001051/1.

% References
\bibliographystyle{aipproc.bst}

\end{document}